# Assessing the feasibility and consequences of nuclear georeactors in Earth's core-mantle boundary region

R.J. de Meijer[a,b,*] and W. van Westrenen[c]

a) Stichting EARTH, de Weehorst, 9321 XS Peize, the Netherlands

b) Department of Physics, University of the Western Cape, Private Bag X17, Bellville 7535, South Africa

c) Faculty of Earth and Life Sciences, Vrije Universiteit Amsterdam, De Boelelaan 1085, 1081 HV Amsterdam, the Netherlands

*Corresponding author, e-mail rmeijer@geoneutrino.nl

**Revised version 22 April 2008**

## Abstract

We assess the likelihood and geochemical consequences of the presence of nuclear georeactors in the core-mantle boundary region (CMB) between Earth's silicate mantle and metallic core. Current geochemical models for the Earth's interior predict that U and Th in the CMB are concentrated exclusively in the mineral calcium silicate perovskite (CaPv), leading to predicted concentration levels of approximately 12 ppm U + Th, 4.5 Ga ago if CaPv is distributed evenly throughout the CMB. Assuming a similar behaviour for primordial [244]Pu provides a considerable flux of neutrons from spontaneous fission. We show that an additional concentration factor of only an order of magnitude is required to both ignite and maintain self-sustaining georeactors based on fast fission. Continuously operating georeactors with a power of 5 TW (terawatt: 1 TW=10[12] W) can explain the observed isotopic compositions of helium and xenon in the Earth's mantle. Our hypothesis requires the presence of elevated concentrations of U and Th in the CMB, and is amenable to testing by direction-sensitive geoneutrino tomography.

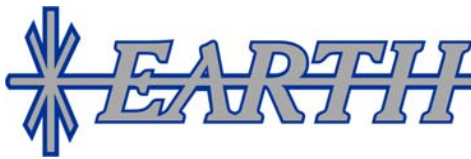

37
38
39
40
41
42
43
44
45

46
47
48
49
50
51
52
53

54
55
56
57
58
59
60
61

62
63
64
65
66
67
68
69
70

71
72
73
74
75
76
77
78
79

# 1. Introduction

Accurate knowledge of the concentrations and distribution of the main heat-producing elements uranium, thorium and potassium in the interior of the Earth is required to constrain our planet's thermal evolution and dynamics. Classically, these elements are considered to be concentrated predominantly in Earth's crust[1], with the underlying mantle and core containing significantly lower concentrations[2]. However, recent measurements of small but significant excesses of $^{142}Nd$ (formed by α-decay of $^{146}Sm$, $t_{1/2}$ = 103 Ma) in terrestrial rocks compared to undifferentiated meteorites[3,4] have dramatically altered this notion.

These $^{142}Nd$ excesses suggest that all terrestrial samples formed from a reservoir with an initial Sm/Nd ratio higher than that in the meteorites that likely formed the building blocks of Earth. This in turn requires the presence of an ancient (> 4.52 Ga) isolated geochemical reservoir in the Earth's interior that has a complementary lower initial Sm/Nd ratio. Mass-balance calculations show that this early, enriched hidden reservoir must also be enriched in other lithophile elements, including K, U and Th. It is now thought that up to 40% of the mantle's inventory of U and Th (equivalent to 20% of the total terrestrial budget) resides in this isolated reservoir[5].

Seismic observations show that the Earth's mantle convects[6], and numerical models show this convection severely limits the possibility for material to remain completely isolated in the bulk mantle for 4.5 billion years[7]. As a result, the only likely locus of an isolated geochemical reservoir is in the lowermost part of the mantle, the core-mantle boundary (CMB) region directly on top of Earth's metallic liquid core. Material that resides at this lower boundary layer for mantle convection has limited interaction with the convection cells. According to Tolstikhin et al.[5] only 6% of the initial isolated reservoir has been mixed back into the remainder of the mantle since it was formed.

These spectacular findings have prompted our re-examination of the likelihood and consequences of the presence of active, natural nuclear reactors (so-called georeactors) in the CMB region. Georeactors in Earth's metallic core as significant heat-producing entities in its interior have been suggested at various times over the past fifty years[8-13]. Various authors[10-13] have estimated their present-day power to be between 4 and 30 TW, which is significant compared to the observed present-day heat flux at Earth's surface (31-44 TW)[14,15]. Uncertainties in the size of the various Earth's heat sources allow for additional heat sources of 0-15 TW, e.g. by georeactors[16]. Analysis of the recent KamLAND data on antineutrino fluxes also set an upper limit of 18 TW produced by georeactors[17]

Previous studies that hypothesized active georeactors invoked gravitationally-induced concentration of actinides to supercritical values in the metallic core of the Earth, either in the centre[11] or on top of the progressively crystallising solid inner core[13]. To date, these studies have been largely shunned by the Earth-Science community, mostly for geochemical reasons. Although laboratory experiments show it is possible to incorporate uranium in iron-rich metallic melts under highly reducing conditions at high pressures and temperatures[18,19], these conditions also lead to incorporation of significant amounts of trace and even major elements (among them Mg, Ca, and the lanthanides) in the metal phase. This is inconsistent with most geochemical models for the composition of the Earth[2,20],

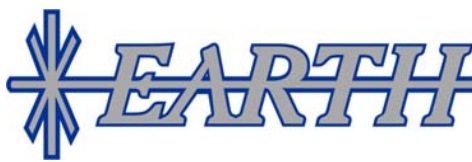

80  which show that Mg, Ca, and the lanthanides cannot be present in the core at concentration
81  levels that would accompany significant U incorporation if undifferentiated meteorites
82  formed the basic building blocks of our planet.

83      In this paper we first combine experimental results quantifying the distribution of U
84  and Th between the main silicate minerals present in Earth's lower mantle, with
85  geochemical estimates for the U and Th content of the CMB region. We also assess the
86  CMB concentration of Pu, because of the potentially important role of its isotope [244]Pu
87  ($t_{1/2}$= 80 Ma) in fission processes in the early Earth. Secondly, we calculate the
88  concentrations of U, Th, and Pu that are required for a georeactor to operate spontaneously
89  and asses the feasibility of U/Th/Pu concentrations in the CMB reaching these levels.

90      After concluding that both starting and sustaining georeactors appears feasible, we
91  analyse some critical geochemical consequences of the presence of georeactors in the
92  CMB. We show that the observed isotopic compositions of Earth's noble-gas inventory for
93  helium, krypton and xenon can be explained through a time-integrated georeactor energy
94  production of $7*10^{29}$ J, equivalent to continuous activity of a single reactor during 4.5 Ga
95  with a power of 5 TW. We also conclude that the fission products of georeactors can be
96  traced in the He and Xe isotopic composition of the Earth's subsurface. Geoneutrino
97  tomography provides the best and, at present, the only way to test the georeactor hypothesis
98  and find an answer to the intriguing question whether nuclear reactors are still operating in
99  the interior of the Earth.

100

101  **Table 1. Masses and isotopic abundances of Th, U and Pu isotopes in the silicate Earth according to**
102  **BSE.**

|  | [232]Th | [235]U | [238]U | [244]Pu | Total mass |
|---|---|---|---|---|---|
| $T_{1/2}$ (Ga) | 14.05 | 0.70 | 4.47 | 0.08 | |
| m ($10^{17}$kg) (t=0) | 3.15 | $5.87*10^{-3}$ | 0.80 | - | 3.95 |
| Isotopic abundance (t=0) | 100% | 0.73% | 99.27% | - | |
| m ($10^{17}$kg).(t=-4.55 Ga) | 3,94 | 0.52 | 1.62 | $1.2*10^{-2}$ | 6.07 |
| Isotopic abundance (t=-4.55 Ga) | 100% | 24.3 | 75.7% | 100 | |

103
104
105      ## 2. Constraints on U, Th, and Pu distribution in the CMB

106      As mentioned in the introduction, current models of Early Earth evolution, based on
107  geochemical analyses of terrestrial material and meteorites, require the presence of a
108  'hidden', deep-Earth geochemical reservoir that remained isolated from convective
109  processes in Earth's mantle for > 4.52 billion years[3]. Tolstikhin et al.[5] state that this
110  reservoir, necessarily located in the CMB, contains up to 20% of the total terrestrial budget
111  of radiogenic heat-producing elements K, U and Th. We assume that Pu behaves similarly,
112  and that the Pu concentration in the BSE equalled 0.19 ppb at 50 Ma after the beginning of
113  the solar system[21]. Table 1 presents the masses of [232]Th, [235]U, [238]U and [244]Pu present in the
114  bulk silicate Earth (BSE, equivalent to continental crust + mantle) today and 4.55 Ga ago,
115  using the BSE model of McDonough[2]. Using the BSE mass of $4.0*10^{24}$ kg, and CMB mass of approximately $2.0*10^{23}$ kg, the concentrations of U and Th in the CMB are calculated to be 80 and 320 ppb at present, respectively. When the isolated reservoir was formed, these values (corrected for nuclear decay) were 210 ppb U, 400 ppb Th and 0.8 ppb Pu. It should be noted that these concentrations depend inversely on the assumed mass of the CMB, which is not well-constrained.

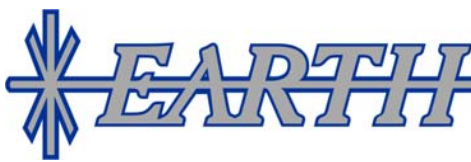

These concentrations are average concentrations for the CMB as a whole. However, high-pressure experiments on putative mantle rocks, in combination with geophysical observations of the density structure of Earth's interior, show that virtually all of Earth's lower mantle (at depths > 660 km) consists of three main minerals: 65 vol% magnesium silicate perovskite (MgPv, nominal chemical composition $(Mg,Fe)SiO_3$), 30 vol% ferropericlase (FP, nominal composition $(Mg,Fe)O$) and 5 vol% calcium silicate perovskite (CaPv, nominal composition $CaSiO_3$).

Experimental studies have also quantified how U and Th distribute themselves between these phases at high pressures and temperatures[22]. Although at present it is impossible to obtain distribution data at CMB conditions (i.e. pressures of approximately 125 GPa and temperatures of 2500-4000 K), experiments at 25 GPa and 2300 °C indicate that U and Th concentrations in calcium silicate perovskite are 3-4 orders of magnitude higher than concentrations in co-existing MgPv[22]. The recently discovered new high-pressure form of MgPv, named postperovskite[23], which may be stable in the CMB, is unlikely to affect this distribution. Ferropericlase in turn generally incorporates even lower concentrations of trace elements than MgPv[24]. No experimental measurements of Pu distribution behaviour are available, but trends in distribution behaviour as a function of ionic radius suggest that Pu favours CaPv over the other lower mantle minerals by 2 orders of magnitude.

Combining this information and assuming 5 wt% of the fully-mixed CMB consists of CaPv, we conclude that upon formation of the isolated reservoir, concentrations of U, Th and Pu in CaPv were 4.3 ppm, 7.9 ppm and 23 ppb, respectively. Areas with lower concentrations of CaPv and/or higher concentrations of U, Th, and Pu may have even higher concentrations of these fissionable nuclei.

### 3. Concentration requirements for the initiation and operation of georeactors

In this section we consider the conditions required for initiation and operation of georeactors in the CMB 4.5 Ga ago. Fission of heavy nuclei like thorium and uranium produces approximately 200 MeV ($3.2*10^{-11}$J) in energy and releases ~three fast neutrons. Several nuclei fission spontaneously; for nuclei like $^{235}U$, $^{238}U$, and $^{244}Pu$ the probability is very small compared to α-decay, but not insignificant for initiating georeactors. The initiation is related to the question if the criticality conditions for a georeactor can be met, while operating conditions are related to the availability of sufficient fuel. Criticality is achieved if the neutron-multiplication factor $k_{eff}$ exceeds unity.

Fission simulations by Ravnik and Jeraj[25] show that a local concentration of U of 1 wt% is required to initiate a georeactor 4.5 Ga ago, and that $k_{eff}$ can not exceed unity for Th/U ratios larger than 1.1. In their calculations, which focused on the possibility for present-day reactors, the availability of Pu was not taken into account. The ratio between spontaneous fission (sf) and α-decay of $^{244}Pu$ is a million times larger than of $^{238}U$. So despite its three orders of magnitude lower concentration than $^{238}U$, spontaneous fission of

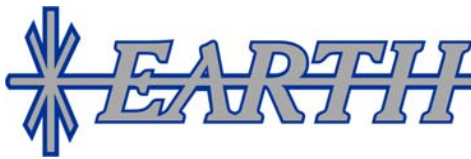

161 $^{244}$Pu, if present, is a more important source for neutrons than spontaneous fission of $^{238}$U
162 by about three orders of magnitude.

163       Every neutron may start a chain reaction, but as long $k_{eff} <1$, all chains will
164 extinguish and the reactor remains subcritical. Nevertheless, under these conditions
165 fissionable materials such as $^{233}$U, $^{239, 240}$Pu are produced and under the conditions sketched
166 above it is likely that a georeactor will become critical, first at low power density, but with
167 increasing fuel production or concentration in geological processes, the power density may
168 increase.

169       Our calculations show that the presence of $^{244}$Pu in the early Earth plays a crucial
170 role in raising $k_{eff}$ to supercritical levels. In a certain mass or volume the value of k is the
171 difference between the number of neutrons released by fission events and the numbers lost
172 due to absorption. Power production is directly related to the production rate of neutrons,
173 which in turn is related to the specific spontaneous fission rate of nuclei. At t=-4.5 Ga ago
174 the mass fractions of $^{235}$U and $^{238}$U were 0.24 and 0.76, respectively (see Table 1). For U
175 this leads to a fission rate of $8.6*10^{-17}$ per atom per year and of $2.8*10^{-11}$ per atom per year
176 for $^{244}$Pu. A concentration of 1% U corresponds to $3.6*10^{-18}*N_A$ kg$^{-1}$.a$^{-1}$ with $N_A$ being
177 Avogadro's number. Interestingly, the fission rate for 4.3 ppm U and 19 ppb Pu (estimated
178 concentrations for CaPv in the CMB) amounts to $2.1*10^{-18}*N_A$ kg$^{-1}$.a$^{-1}$, very close to the
179 value for the requirement for criticality in an environment without Th. The presence of Th
180 reduces the value of $k_{eff}$ to 0.8 at a Th/U ratio of 1.8 in absence of $^{244}$Pu[25]. A local
181 concentration factor for Pu in the CMB of only an order of magnitude (compared to a
182 homogeneous distribution of CaPv) is required to sufficiently raise $k_{eff}$ and initiate a
183 georeactor. Seismic observations show that at present the CMB region is incredibly
184 heterogeneous with respect to seismic wave velocities[26]. Volumes exhibiting both higher-
185 than-average and lower-than-average wave propagation speeds with diameters as small as
186 30 km are now resolvable. Although the precise nature of these heterogeneities remains
187 unclear, we suggest that significant local concentration factors appear possible even today.
188 The dynamics of the CMB 4.5 Ga ago are largely unstudied. The higher rotation rate of
189 Earth at that time, in conjunction with higher temperatures, will likely have facilitated local
190 concentration of density heterogeneities to levels that exceed those currently observed, due
191 to centrifugal forces and buoyancy effects related to local heating,.

192       Assuming that one or more georeactors were initiated in the CMB shortly after the
193 formation of the isolated reservoir, we now consider the conditions that need to be met for a
194 reactor to continue operation. In the likely absence of considerable amounts of hydrogen in
195 the CMB, neutrons produced in fission will only gradually be thermalised and hence have a
196 mean free path of the order of a hundred metres. Apart from introducing fission, they will
197 be mainly captured by $^{238}$U and $^{232}$Th, producing $^{239}$Pu and $^{233}$U, respectively, which are in
198 turn fissionable materials. In addition, $^{239}$Pu decays by α-particles to $^{235}$U. So although
199 especially Th may be hampering the initiation of a georeactor, the capture will eventually
200 lead to an increase in fuel for a potential georeactor. The potential georeactor after initiation
201 will run as a breeder reactor and will be self-sustainable due to generation of its own fuel
202 by the processes mentioned above[10,27,28].

203       The question remaining is whether the U and Th concentrations in CaPv in the
204 CMB of 4.3 and 7.9 ppm are sufficient for sustainable georeactors 4.5 Ga ago (note that the
205 19 ppb of $^{244}$Pu does not play a role here). To maintain a nuclear reaction, $k_{eff}$ = 1, one of

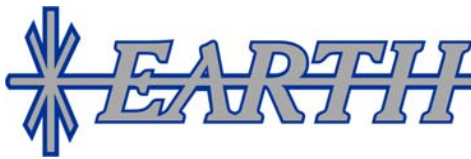

206 the three neutrons produced in fission must lead to a new fission process; the two others
207 should be absorbed by the CMB materials. This leads to the following simplified relation
208 between the masses of fissionable, $m_f$, and non-fissionable material, $m_r$:

209
$$\frac{m_f}{m_r} = 0.5 \frac{\sigma_a}{\sigma_f} * \frac{mol_r}{mol_f},$$

210 where $\sigma_a$ and $\sigma_f$ are the atomic cross sections for neutron absorption in the non-fissionable
211 material and the cross section for fission, respectively, and $mol_r$ and $mol_f$ represent the
212 average molecular weights. According to Rinard[29] the cross-section ratio at 1 MeV neutron
213 energy is about 2000-3000, indicating that the homogeneously mixed CaPv and a
214 homogeneous mixture of U and Th constitute concentrations which are a factor of twenty
215 too low for the assumed mass of the CMB. Such an additional concentration factor is not
216 unlikely for a dynamic system such as the CMB. During local partial melting, for example,
217 concentration factors of several orders of magnitude could easily occur. In the case of CaPv
218 cyclic motions due to radiogenic heating, followed by cooling could have been caused by
219 the interplay of gravitational and thermal stabilities. In an operating georeactor, such a
220 mechanism could 'clear' the reactor from its less dense fission products.
221     A concentration factor of twenty is even desired to avoid the total CMB becoming a
222 georeactor. We would like to point out that independently the concentration factors for
223 ignition and sustainable operation come out to be comparable.
224
225             **4. Geochemical consequences of georeactors**
226     After concluding that georeactors in the CMB are feasible, we examine some of
227 their geochemical consequences. We limit ourselves to a first-order approximation, which
228 implies (1) Binary fission of $^{232}$Th, $^{235}$U and $^{238}$U only will be considered (no Pu fission is
229 incorporated); (2) Initial concentrations remain over the entire period of activity; (3)
230 Thermal and fast fission contribute equally; (4) All three fuel nuclei have the same
231 probability to fission; (5) The production probabilities of the nuclides are taken from the
232 code SCALE[30]; (6) The fission products are assumed to mix homogeneously with the
233 material in the CMB; (7) A fraction of the fission products (10%) enters the overlying
234 mantle and is homogeneously mixed by mantle convection.
235     We concentrate on the noble gases helium (section 4.1), krypton and xenon (section
236 4.2). Neon and argon are not taken into account because they are hardly produced in binary
237 fission. In addition argon consists mainly of $^{40}$Ar, a decay product of $^{40}$K. We also present
238 calculations for Se, Mo, Ru and Pd (section 4.3), elements that could be of interest in
239 testing the georeactor hypothesis because of their low abundance in the silicate Earth,
240 combined with high probabilities for production of some of their isotopes by binary fission.
241
242 *4.1 Helium*
243     The $^3$He/$^4$He atomic ratio in Earth's atmosphere[31] is $1.37*10^{-6}$. In the radioactive
244 decay series of uranium and thorium α-particles are emitted that become $^4$He atoms. In
245 spontaneous fission of $^{235}$U and $^{238}$U $^3$H (tritium) is produced which decays with a half-life
246 of 12.3 years to $^3$He. For our considerations this conversion takes place instantaneously and
247 with 100% efficiency. Table 2 presents the half-life for radioactive decay of the Th and U
248 nuclides, the number of α-particles produced in the decay, the half-life for spontaneous fission, the relative probability of fission in the decay, the probabilities per fission for the formation[30] of $^3$H and $^4$He, and the expected $^3$He/$^4$He ratios. In this ratio the $^3$He value is the production probability via spontaneous fission only. For $^4$He, the sum of production in natural decay and in spontaneous fission is used. From the last column of Table 2 it is obvious that the $^3$He/$^4$He ratio in the atmosphere cannot be explained solely by natural decay of U, Th and Pu including spontaneous fission. Taking into account the decayed masses of the various nuclides a ratio of $6.6*10^{-10}$ is obtained.

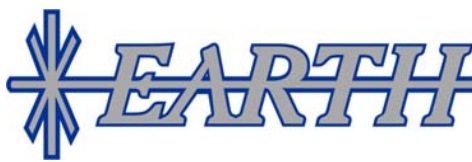

**Table 2. Decay and fission production of $^3$H and $^4$He probability for U and Th.**

| nuclide | $t_{1/2}$(Ga) | # α | $t_{1/2}$-sf (Ga) | $P_{rel}$(sf) | $P_{sf}$($^3$H) | $P_{sf}$($^4$He) | $^3$He/$^4$He |
|---|---|---|---|---|---|---|---|
| $^{232}$Th | 14.1 | 6 | >$10^{12}$ | <$10^{-13}$ | 0 | 0 | <$1*10^{-18}$ |
| $^{235}$U | 0.71 | 6 | $1.9*10^8$ | $0.37*10^{-8}$ | $1.00*10^{-4}$ | $1.88*10^{-3}$ | $0.62*10^{-13}$ |
| $^{238}$U | 4.51 | 8 | $6.2*10^6$ | $0.73*10^{-6}$ | $1.06*10^{-4}$ | $1.62*10^{-3}$ | $0.97*10^{-11}$ |
| $^{244}$Pu | 0.08 | 3 | 25 | 0.0032 | $1.41*10^{-4}$ | $1.94*10^{-3}$ | $1.50*10^{-7}$ |

In Table 3 the probabilities for thermal and fast fission are given together with the $^3$He/$^4$He ratios for pure fission. By comparing the last columns in Tables 2 and 3, it is obvious that fission in a georeactor, followed by transport of He to the atmosphere, would lead to a significant change in the atmospheric $^3$He/$^4$He ratio.

**Table 3. Production probabilities per fission and $^3$He/$^4$He ratios for thermal (th) and fast (ff) fission.**

| nuclide | $P_{th}$($^3$H) | $P_{th}$($^4$He) | ($^3$He/$^4$He)$_{th}$ | $P_{ff}$($^3$H) | $P_{ff}$($^4$He) | ($^3$He/$^4$He)$_{ff}$ |
|---|---|---|---|---|---|---|
| $^{232}$Th | $1.01*10^{-4}$ | $2.40*10^{-3}$ | $0.42*10^{-1}$ | $1.01*10^{-4}$ | $2.40*10^{-3}$ | $0.42*10^{-1}$ |
| $^{235}$U | $1.00*10^{-4}$ | $1.88*10^{-3}$ | $0.53*10^{-1}$ | $1.44*10^{-4}$ | $2.06*10^{-3}$ | $0.70*10^{-1}$ |
| $^{238}$U | $1.06*10^{-4}$ | $1.62*10^{-3}$ | $0.65*10^{-1}$ | $1.06*10^{-4}$ | $1.62*10^{-3}$ | $0.65*10^{-1}$ |

Next we estimate the $^3$He and $^4$He production in the CMB due to radioactive decay and fission processes and the $^4$He production in the remaining mantle. For the remaining mantle the U and Th content is calculated from the concentration in the upper mantle according to the BSE model (0.0065 ppm U and 0.0173 ppm Th). No primordial concentrations of He are assumed to be present.

Table 4 presents the masses of Th and U 4.55 Ga ago and presently in the CMB and the remaining mantle, the changes in Th and U masses due to natural decay and the resulting production of $^4$He for the two compartments. In this and the following comparisons of isotopic ratios, Pu does not play a significant role due to its relative small mass and has been left out. The amounts of $^4$He produced by spontaneous fission can be neglected and have not been included. The values in the two last columns of Table 4 are given by:

$$m(^4He) = \frac{4}{A} N_\alpha \Delta m \qquad (1).$$

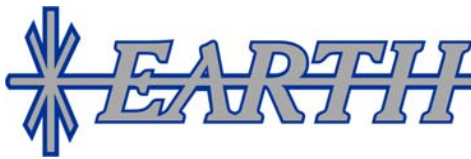

279 In this equation the masses m and Δm are in kilograms, $N_\alpha$ is the number of α-particles
280 produced in the decay chain per nucleus and A is the atomic number of the fissioning
281 nucleus.
282

283 **Table 4. Masses of U and Th at present and 4.55 Ga ago, the masses lost due to decay and the number**
284 **of $^4$He atoms produced during the last 4.55 Ga in the CMB and the remaining Mantle (RM).**

| nuclide | Present mass CMB ($10^{16}$ kg) | Mass at t=-4.55 Ga ($10^{16}$ kg) | $\Delta m_{CMB}$ ($10^{16}$ kg) | Present mass RM ($10^{16}$ kg) | Mass at t=-4.55 Ga ($10^{16}$ kg) | $\Delta m_{RM}$ ($10^{16}$ kg) | $^4He_{CMB}$ (*$10^{15}$kg) | $^4He_{RM}$ (*$10^{15}$kg) |
|---|---|---|---|---|---|---|---|---|
| $^{232}$Th | 6.30 | 7.88 | 1.58 | 6.92 | 8.65 | 1.73 | 1.63 | 1.79 |
| $^{235}$U | 1.17 | 1.04 | 1.02 | 1.90 | 1.68 | 1.66 | 1.04 | 1.69 |
| $^{238}$U | 1.60 | 3.24 | 1.64 | 2.59 | 5.24 | 2.65 | 2.21 | 3.56 |
| Total | | | | | | | 4.88 | 7.04 |

285
286 To assess the consequences of a georeactor we assume that the georeactor has
287 produced an energy of 7.1*$10^{29}$J, corresponding to continuously operating at a power of 5
288 TW for the last 4.5 Ga. Since one fission event produces about 200 MeV or 3.2*$10^{-11}$J, one
289 may calculate that at the U and Th composition 4.5 Ga ago 8.2*$10^{13}$J was produced by the
290 fission of 1 kg mixture. Hence 8.6*$10^{15}$kg of mixture fissioned in the georeactor.
291

292 **Table 5. Distribution of the fissioned mass for the three radionuclides and the resulting $^3$He production**
293 **for operating a 5TW georeactor for 4.5 Ga.**

| nuclide | fissioned mass (*$10^{15}$ kg) | $^3$He (*$10^9$ kg) |
|---|---|---|
| $^{232}$Th$_{th}$ | 2.80 | 3.64 |
| $^{235}$U$_{th}$ | 0.37 | 0.47 |
| $^{238}$U$_{th}$ | 1.15 | 1.53 |
| $^{232}$Th$_{ff}$ | 2.80 | 3.64 |
| $^{235}$U$_{ff}$ | 0.37 | 0.67 |
| $^{238}$U$_{ff}$ | 1.15 | 1.53 |
| Total | 8.6 | 11.5 |

294
295 Assuming all helium, produced either by natural decay or by fission, remains inside
296 the CMB the combination of the data in Tables 4 and 5 leads to $^3$He/$^4$He atomic ratio of
297 3.14*$10^{-6}$. In case the $^3$He combines with all the $^4$He in the mantle and is distributed
298 homogeneously an atomic ratio of 1.28*$10^{-6}$ is obtained. Higher ratios are obtainable by
299 assuming inhomogeneities, a larger time-integrated power delivered by georeactors or a
300 smaller U and Th content of the CMB.
301

302 *4.2 Krypton and xenon*
303 Fission in general will lead to a preferential population of neutron-rich isotopes of
304 elements. An additional consequence of georeactors may be observed in the isotope ratios
305 of elements with a low abundance but a high probability for being formed in fission. In this
306 section we discuss krypton and xenon, which may play an essential role in testing a CMB-
307 georeactor hypothesis. For each of these fission chains the probability of fission is obtained as the sum of the probabilities for all isotones decaying to a particular isotope of an element. For example in the formation of [81]Kr, the sum extends over the production of the nuclei [81]Kr, [81]Br, [81]Se, [81]As, [81]Ge and [81]Ga, but for [80]Kr only the production of [80]Kr and [80]Br contribute since [80]Se is stable.

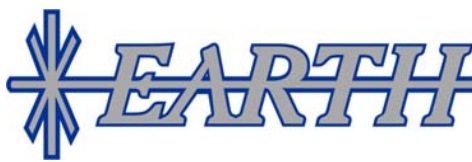

To estimate the possible modifications in isotope ratios for Kr and Xe, we have to know the natural abundances of these two elements in various compartments of the Earth. In our atmosphere, the abundances for Kr and Xe are well-established as 1.14 ppm and 87 ppb by volume, respectively. To estimate the abundances of these elements in the mantle and the CMB we take the abundances in C-chondrites according to Lodders[32]. Lodders[32] indicates that these abundances may be altered due to outgassing. The uncertainty is especially large for the lighter noble gases. The chondritic values of $5.22*10^{-11}$ and $1.74*10^{-10}$ for Kr and Xe, respectively, compared to the atmospheric concentrations, indicate that apparently outgassing of Kr still is of importance. We start from the chondritic value for xenon (i.e. assuming no significant outgassing for this heavy noble gas), and scale the abundance of krypton in proportion to the atmospheric ratio. Implicitly this assumes that the outgassing of the meteorites at low temperatures over 4.55 Ga is of the same order of magnitude as the outgassing on Earth at high temperatures in the first 0.1 Ga. This results in a corrected Kr abundance of $2.28*10^{-9}$ in both the mantle and the CMB, leading to calculated masses of Kr in the CMB and the mantle of $5.4*10^{14}$ and $9.1*10^{15}$ kg, respectively. Moreover we assume that the isotopic ratios in the primordial Kr and Xe are identical to the present values in air.

**Table 6. Production (kg) of various stable isotopes of Kr and their isotopic abundances in the CMB and the Mantle due to the continuous 5TW georeactor.**

| Krypton Isotope | Mass due to fission (kg) | Rel. abund (%) | Air. rel. Abund. (%) | Nat +fission in CMB; rel. abund. (%) | Full mantle mixing;rel. abund. (%) | 10% mixing; rel.abund. (%) |
|---|---|---|---|---|---|---|
| 78 | $1.33*10^4$ | <0.1 | 0.35 | 0.27 | 0.35 | 0.35 |
| 80 | $4.26*10^8$ | <0.1 | 2.25 | 1.73 | 2.22 | 2.25 |
| 82 | $1.12*10^{11}$ | 0.08 | 11.6 | 8.93 | 11.4 | 11.6 |
| 83 | $2.47*10^{13}$ | 18 | 11.5 | 13.0 | 11.6 | 11.5 |
| 84 | $3.90*10^{13}$ | 28 | 57.0 | 50.4 | 56.6 | 57.0 |
| 86 | $7.38*10^{13}$ | 54 | 17.3 | 25.7 | 17.8 | 17.4 |

Table 6 presents the mass of the various stable isotopes of krypton and the derived isotopic abundances. In column 4 the ratios as present in our atmosphere have been presented. In the remaining columns data for a number of scenarios in which the fission product krypton is fully mixed with the CMB (column 5) and the Mantle (column 6), as well as a 10% mixture of the fission products with the mantle (column 7) are given as isotopic ratios. From table 6 one can observe that the isotopic abundances in the fission products are strongly shifted towards the heavier isotopes. Due to mixing with the naturally present Kr the effect is already strongly diminished in the CMB and leads to a hardly noticeable effect in case of full mixing with the mantle and no noticeable effect if the hidden reservoir is mixed for only 10% with the entire mantle.

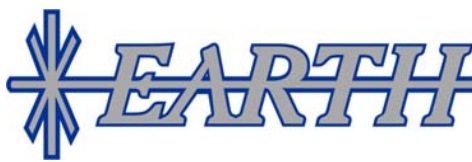

A similar approach has been taken for xenon. From the chondritic abundance[32] of $1.74*10^{-10}$ the mass of xenon in the CMB and mantle are $4.14*10^{13}$ and $6.96*10^{14}$ kg, respectively. In Table 7 the data are presented in a similar way as table 6. From a comparison between rows 3 and 4 one notices again a shift of the abundances towards the heavier isotopes. Contrary to Kr where the heaviest isotope is most strongly produced, in Xe the maximum occurs at the one but heaviest isotope ($^{134}$Xe). Since on the one hand the production of xenon in fission is stronger for krypton and on the other hand the natural elemental abundances for xenon are lower than for krypton, fission products have a larger influence on the isotopic ratios after mixture. Hence the changes in isotopic composition of mantle-derived xenon gasses are expected to be more outspoken than for krypton. Another aspect that can be noticed is the "trapping" of fission products by stable Te isotopes ($^{128}$Te and $^{130}$Te). This trapping boosts the relative abundance of $^{129}$Xe relative to its neighbours.

**Table 7. Production (kg) of various stable isotopes of Xe and their isotopic abundances in the CMB and the Mantle due to the continuous 5 TW georeactor.**

| Xenon Isotope | Mass due to fission (kg) | Rel. abund (%) | Air. rel. Abund. (%) | Nat +fission in CMB; rel. abund. (%) | Full mantle mixing;rel. abund. (%) | 10% mixing; rel.abund. (%) |
|---|---|---|---|---|---|---|
| 128 | 0 | 0 | 1.91 | <0.1 | 0.76 | 1.66 |
| 129 | $6.55*10^{13}$ | 6.2 | 26.4 | 6.95 | 14.2 | 23.7 |
| 130 | $1.39*10^{11}$ | <0.1 | 4.1 | 0.14 | 1.64 | 3.6 |
| 131 | $1.44*10^{14}$ | 13.6 | 21.2 | 13.8 | 16.6 | 20.2 |
| 132 | $2.33*10^{14}$ | 22.1 | 26.9 | 22.3 | 24.0 | 26.3 |
| 134 | $3.27*10^{14}$ | 31.0 | 10.4 | 30.3 | 22.8 | 13.1 |
| 136 | $2.86*10^{14}$ | 27.1 | 8.9 | 26.5 | 19.9 | 11.3 |

### 4.3 Other elements

In this section we discuss the elements Se, Mo, Ru and Pd, selected on the basis of their relatively low abundances (75, 50, 5 and 4 ppb, respectively) in the Earth's silicate mantle, and their relative large probability to be formed in fission.

In the case of Se, although the isotopic ratios in the fission products deviate strongly from the natural ratios, the mass of natural selenium in the CMB is compared to the fission production about two orders of magnitude larger for each isotope. Hence, there are only minute changes (<0.1%) in the mixed CMB material.

The production cross-sections in fission for Mo isotopes are considerably larger than for selenium, so that one expects a larger effect of georeactors on the mantle molybdenum isotopic composition. Again, fission production favours the production of heavier isotopes, but the difference with the natural abundances is not as large as for Kr, Xe and Se. This feature significantly reduces the signal of a possible 5 TW georeactor in the CMB. Full mixing of the fission products with the entire mantle only leaves small effects

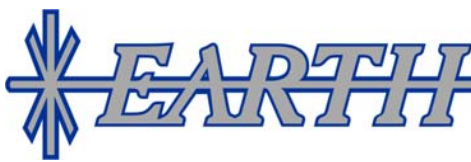

373 (<1%) and a 10% mixing of the fission products with the mantle leaves an even smaller
374 effect (<0.1%).
375       Palladium (Pd) and ruthenium (Ru) are elements with a low abundance in the BSE
376 (3.9 and 5.0 ppb, respectively). They are therefore likely candidates in which the
377 consequences of a possible georeactor in the CMB will be clearly noticeable. For Ru, in
378 contrast to the previous elements the highest fission yields are not obtained for the heaviest
379 isotopes, but similar to the natural ratios in the middle group. One of the consequences is
380 that mixing with CMB or mantle material, the differences between the isotopic abundances
381 in the mixtures and in the BSE are rapidly diminishing. With a leak of 10% of the fission
382 products into a well-mixed mantle, the differences become less than 1%.
383       For Pd, the highest fission yield is obtained for $^{105}$Pd and $^{106}$Pd, isotopes that also
384 have a high natural abundance. One of the consequences is that at mixing with CMB or
385 mantle material the differences in abundance for the mixtures and for the BSE are rapidly
386 diminishing. Moreover, the fission yields are one to two orders of magnitude smaller than
387 for e.g. molybdenum. The differences completely disappear (< 0.1%) when mixing with the
388 overlying mantle is included.
389
390 **5. Assessment of the noble gases**
391 *5.1 Helium*
392       The measured abundance of $^{3}$He relative to $^{4}$He in terrestrial reservoirs forms a
393 stringent test for the georeactor hypothesis. Appendix 1 presents the observed $^{3}$He/$^{4}$He
394 ratios in various terrestrial compartments and tries to identify the main sources for the two
395 helium isotopes. The natural branch of U and Pu has a small branch of spontaneous fission,
396 resulting in a fixed $^{3}$He/$^{4}$He ratio < $10^{-9}$; several orders of magnitude smaller than found
397 anywhere on Earth. This perceived excess in $^{3}$He has been termed the helium paradox[33,34].
398       Primordial $^{3}$He as a significant source for this 'excess' is unlikely due to the high
399 temperatures in the early ages of the Earth which caused elements far less volatile than He
400 (e.g. Na, K, and Cl) to be partially removed from Earth's mantle. The fact that presently the
401 $^{3}$He concentrations in wells and plumes range over several orders of magnitude[34] argues
402 against well-mixed primordial $^{3}$He as a source as well. As detailed in section 4.1, the
403 continuous presence of a 5 TW georeactors in the CMB leads to $^{3}$He/$^{4}$He ratios of $3.14*10^{-6}$
404 for a homogeneously mixed CMB, and $1.28*10^{-6}$ for a well-mixed mantle. These estimates
405 of $^{3}$He/$^{4}$He ratios are based on an even distribution of the helium isotopes in the various
406 compartments of the Earth's mantle. Isolated georeactors would lead to an expected
407 $^{3}$He/$^{4}$He ratio of $0.5-1*10^{-1}$ (Table 3). As georeactors are necessarily localised, a range of
408 ratios between $1.28*10^{-6}$ and $1*10^{-1}$ is expected for samples coming from the deep Earth.
409 Observed values in deep wells and mantle plumes in fact range from 1-50 times the
410 atmospheric value of $1.37*10^{-6}$, easily within the range obtainable from localised
411 georeactors. Georeactors thus provide a natural source of $^{3}$He, quantitatively resolving the
412 He paradox.
413
414 *5.2 Krypton and xenon*
415       The Kr isotopic composition of mantle samples is typically identical to that of the
416 atmosphere. Spontaneous fission of $^{238}$U leads to minor production of $^{83,84,85}$Kr, but these
417 fission products have not been detected in basalts to date because of the small yields in a mass region where Kr nuclides are relatively abundant. From our calculations no significant deviations from the atmospheric ratios are expected in mantle material, in good agreement with the data.

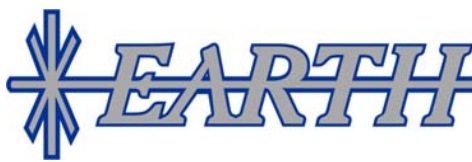

On the other hand, the xenon isotopic composition of samples from Earth's mantle differs significantly from that of Earth's atmosphere[35]. Most notably, excesses in $^{129}$Xe and $^{136}$Xe in mantle rocks have been identified[36,37]. Our current atmosphere has a $^{129}$Xe/$^{130}$Xe ratio of 6.48, and a $^{136}$Xe/$^{130}$Xe ratio of 2.17, but estimates for the current mantle source for mid-ocean ridge basalts show $^{129}$Xe/$^{130}$Xe of 8.2 and $^{136}$Xe/$^{130}$Xe of 2.7.

$^{129}$Xe excesses have been interpreted as being derived from extinct radioactivity of $^{129}$I, while excesses in $^{136}$Xe are supposedly derived from spontaneous fission of $^{238}$U and $^{244}$Pu. From Table 7 it can be deduced that $^{129}$Xe/$^{130}$Xe and $^{136}$Xe/$^{130}$ ratios are 8.7 and 12.1 for a complete mixture of the fission products with the mantle, decreasing to 6.7 and 3.2 for a 10% mixture of georeactor fission products in the mantle.

The latter is in surprisingly close agreement with the observations. Closer agreement could be achieved easily when considering some of the assumptions in our first-order model. In particular, fast and thermal fission are assumed to contribute equally to the fission processes, and the Xe concentrations of the mantle are assumed to be equal to the concentrations found in chondritic meteorites[32]: $1.74*10^{-10}$. These assumptions have a significant effect on the predicted xenon isotopic ratios.

Our calculations show that due to the trapping by $^{130}$Te, $^{130}$Xe in the CMB is dominated by three orders of magnitude by the chondritic concentrations. For $^{129}$Xe in the CMB, the mass from fission is an order of magnitude larger than the chondritic mass, and the contributions from Th and U are 0.8 and 0.2, respectively. For $^{136}$Xe the mass in the CMB due to fission is two orders of magnitude larger than the chondritic mass. However, the contribution of U has increased to one third. For a mixture of 10% of the fission-produced Xe-isotopes with those present in the mantle according to the chondritic abundances, the contribution of fission to the $^{129}$Xe isotope mass is only 3%, but for $^{136}$Xe the contribution is still 30%.

In summary, minor changes of the contributions would lead to perfect agreement between predictions from georeactor products and observed xenon isotopic ratios in Earth's mantle. Although we feel that such tweaking is not justified in view of our crude, first-order approximation, we conclude that the Xe isotopic record of mantle samples is consistent with the georeactors hypothesis.

## 6. Proposed test of the georeactor hypothesis

Our model assumes that areas with elevated concentrations of U and Th continue to be present within the CMB even today. Such locations produce heat by either natural decay or possibly as georeactors[10,27]. Both in natural decay and in fission, antineutrinos will be produced that reach the Earth's surface without any noticeable absorption. The two processes can be distinguished by the antineutrino energy spectrum. In case of natural decay the antineutrinos of U and Th have distinctly different energies up to 3 MeV. In case of fission the antineutrino's have a bell shaped distribution from about 2 MeV up to 8 MeV, similar to the antineutrinos emitted by nuclear-power reactors.

In July 2005 the KamLAND collaboration[38] presented the first evidence for geoneutrinos (antineutrinos produced inside Earth). Development of direction-sensitive

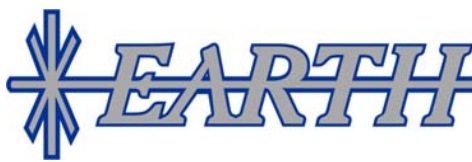

463 antineutrino detectors[39,40] would provide a critical test of our hypothesis. The first step in
464 the development of such type of detector has recently successfully been completed[41,42].
465 Measurements with a network of these detectors could reveal whether elevated
466 concentrations of U and Th are present in the CMB, and if so, identify the process by which
467 heat is produced.
468
## 7. Conclusions

470 Geochemical models for the Earth's interior lead to concentration levels of U + Th
471 in the mineral calcium silicate perovskite present in the CMB of approximately 12 ppm 4.5
472 Ga ago, requiring an additional concentration factor of only ~20 to maintain georeactors
473 based on fast fission. A similar concentration factor is required to ignite the georeactor. The
474 main source of neutrons in that case is supplied by $^{244}$Pu. This nuclide is estimated to be
475 present in calcium perovskites at a concentration of 19 ppb. Continuously operating
476 georeactors with a power of 5 TW and natural abundances based on chondritic and/or BSE
477 values can explain the observed deviations of the isotopic compositions of helium and
478 xenon in the Earth's mantle. For krypton, selenium, molybdenum, palladium and
479 ruthenium, the absence of measured isotopic anomalies in mantle samples is also expected.
480 Our hypothesis can eventually be tested by direction-sensitive geoneutrino tomography.
481

## Acknowledgments

483 The authors would like to thank dr. Jan Leen Kloosterman, Technical University
484 Delft, NL, for providing us with the results of the SCALE calculations and for discussions
485 regarding ignition and sustainability of georeactors. We are indebted to Prof. W.F.
486 McDonough, University of Maryland, USA, and an anonymous reviewer for valuable
487 discussions and suggestions.

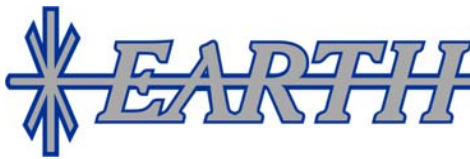

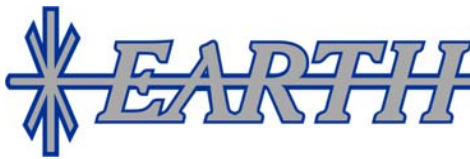

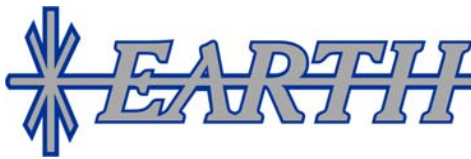

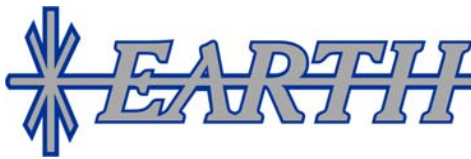

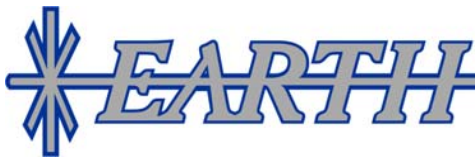

628 Appendix 1: Earth's helium budget
629

630     This appendix addresses the observed $^3$He/$^4$He ratios in various compartments of the
631 Earth and its atmosphere and tries to identify the main sources for these two helium
632 isotopes.

633

634 *Solar wind*

635     The helium content of the solar wind has been measured[43] by analysing metallic
636 glass aliquots flown on the satellite Genesis collecting data for 887 days. The results are a
637 helium particle flux of $9.72*10^6$ cm$^{-2}$ s$^{-1}$ and a $^3$He/$^4$He ratio of $(4.62\pm0.04)*10^{-4}$,
638 corresponding to a fluence of $2.59*10^6$ kg a$^{-1}$ $^4$He and about 900 kg a$^{-1}$ $^3$He at the outer
639 atmosphere. Due to the shielding by the geomagnetic field about 1% enters the Earth's
640 atmosphere.

641

642 *Atmosphere*

643     The Earth's atmosphere has a mass of about $5.1*10^{18}$ kg. The helium content is
644 about $5*10^{-4}$ % by volume, corresponding to a mass of $3.71*10^{12}$ kg. The $^3$He/$^4$He ratio
645 equals[31] $1.37*10^{-6}$. Estimated fluxes in and out of the Earth's atmosphere[44] are tabulated in
646 Table A1.

647

648 **Table A1. Fluxes in and out of the Earth's atmosphere of helium isotopes and corresponding decay and**
649 **net ingrowth rates. (ET= extra-terrestrial; EX= exhalation from the surface).**

|  | Influx$_{ET}$ (kg a$^{-1}$) | Influx$_{EX}$ (kg.a$^{-1}$) | Escape (kg a$^{-1}$) | $\lambda_{esc}$ (a$^{-1}$) | $\lambda_{ing}$ (a$^{-1}$) |
|---|---|---|---|---|---|
| $^3$He | 8.0 | ? | 3.2 | $6.3*10^{-7}$ | $9.4*10^{-7}$ |
| $^4$He | ? | $2.1*10^6$ | $6.4*10^4$ | $1.7*10^{-8}$ | $5.7*10^{-7}$ |

650

651     Table A1 shows that the $^3$He and $^4$He concentrations are not in equilibrium. The
652 question marks indicate unknown values. They could be estimated from either the $^3$He/$^4$He
653 ratio in the solar wind for the extra-terrestrial influx of $^4$He or from the atmospheric ratio
654 for the $^3$He exhalation from the Earth's surface. In both cases the overall picture hardly
655 changes and the discrepancy between influx and escape only increases. Table A1 shows
656 that if the $^3$He/$^4$He ratio in the Earth's continental- and oceanic-crust and sediments is
657 similar to that in the Earth's atmosphere, the $^3$He concentration in the atmosphere is
658 dominated by the extra-terrestrial influx and the $^4$He concentration by the influx from the
659 Earth's surface.

660     In a simple compartment model one can define removal rates of $^3$He and $^4$He. The
661 values in Table A1 correspond to half-life times of ~1 and ~40 Ma, respectively. This
662 difference is considerably larger than expected from the mass difference from the two
663 isotopes. In a similar way the ingrowth can be expressed as a net ingrowth rate. The values
664 in Table A1 indicate that the total mass of $^3$He is likely to increase at a faster rate than $^4$He,
665 but the uncertainties in the numbers are too large to make a firm statement.

666     From these data it can be concluded that, in the atmosphere, the $^3$He concentration is
667 primarily determined by extra-terrestrial input and the $^4$He concentration by exhalation

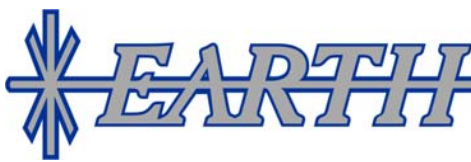

668 from the continental- and oceanic-crusts and sediments. The atmosphere's $^3$He/$^4$He ratio is
669 therefore not a reflection of the $^3$He in the Earth's interior.
670
671 *The continental and oceanic crusts*
672 For the assessment of the helium budget of the Earth's interior we consider a
673 subdivision in upper and lower mantle, and adapt best estimates for the uranium and
674 thorium contents using recent Bulk Silicate Earth (BSE) models. We add to the usual
675 division the CMB, and assume that half of the U and Th of the lower mantle is concentrated
676 in the CMB, roughly in agreement with Tolstikhin et al.[45]. It yields approximately similar
677 values for U and Th concentrations in the upper and lower mantle.
678
679 **Table A2. The present U and Th contents according to BSE of various parts of the Earth's interior and**
680 **the present rates of $^4$He production, due to natural decay.**

| compartment | $^{238}$U (kg) | $^{235}$U (kg) | $^{232}$Th (kg) | $^4$He (kg a$^{-1}$) |
|---|---|---|---|---|
| crust | $3.5*10^{16}$ | $2.6*10^{14}$ | $1.4*10^{17}$ | $1.5*10^6$ |
| Upper mantle | $6.2*10^{15}$ | $4.5*10^{13}$ | $1.7*10^{16}$ | $2.2*10^5$ |
| Lower mantle | $1.9*10^{16}$ | $1.4*10^{14}$ | $7.7*10^{16}$ | $8.0*10^5$ |
| CMB | $1.9*10^{16}$ | $1.4*10^{14}$ | $7.7*10^{16}$ | $8.0*10^5$ |

681
682 Table A2 presents, in addition to the volumes of the upper and lower mantle, the
683 mass of thorium as well as of the two uranium isotopes. From the half-life time, and the
684 fact that each decaying $^{238}$U atom produces 8 atoms of $^4$He and both $^{235}$U and $^{232}$Th produce
685 6 $^4$He atoms each, the production rate of $^4$He in the various compartments follows
686 straightforwardly. From the $^4$He value in the crust it can be concluded that, within the
687 accuracies, this number is not significantly different from the $^4$He exhalation rate at the
688 Earth's surface of $2*10^6$ kg a$^{-1}$. It therefore can be concluded that the crust may be
689 considered as a closed system for $^4$He. This means that we may ignore significant transport
690 of helium from the deeper Earth to the crust.
691
692 *The mantle and the CMB*
693 Without significant transport from the mantle to the crust, $^4$He will build up over
694 time due to natural decay of U and Th. The amount is calculated from the present U and Th
695 content, using the decay constants for radioactive decay to derive the amount of each the
696 uranium isotopes that has decayed in the last 4.5 Ga. Using the number of $^4$He atoms
697 produced in each of the decay chains the total amount of the $^4$He isotope can been
698 calculated.
699
700 **Table A3. The change in mass of $^{238}$U, $^{235}$U and $^{232}$Th due to natural decay and the total amount of $^4$He**
701 **for the three compartments in the mantle over the past 4.5 GPa.**

| compartment | $\Delta^{238}$U (kg) | $\Delta^{235}$U (kg) | $\Delta^{232}$Th (kg) | $^4$He (kg) |
|---|---|---|---|---|
| Upper mantle | $6.3*10^{15}$ | $4.0*10^{15}$ | $4.2*10^{15}$ | $1.7*10^{15}$ |
| Lower mantle | $2.0*10^{16}$ | $1.2*10^{16}$ | $1.9*10^{16}$ | $6.0*10^{15}$ |
| CMB | $2.0*10^{16}$ | $1.2*10^{16}$ | $1.9*10^{16}$ | $6.0*10^{15}$ |

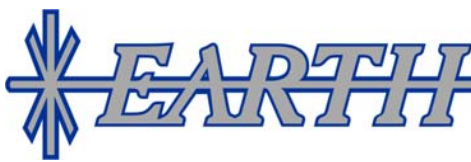

Table A3 lists the mass of uranium isotopes and thorium that has decayed over the last 4.5 Ga. The decay of $^{238}$U has the largest contribution to the $^4$He production, but interestingly enough one notices that despite the small abundance of $^{235}$U at present, this uranium isotope has a larger contribution than $^{232}$Th.